\documentstyle[12pt]{article}
\begin{document}
\baselineskip=24pt
\begin{titlepage}
\vspace{-20ex}
\vspace{1cm}

\centerline{\large \bf Weak Semileptonic Decays of Heavy Baryons}

\vspace{0.3cm}

\centerline{\large \bf Containing Two Heavy Quarks}

\vspace{6.4ex}
\centerline{Xin-Heng Guo$^{1,2}$, Hong-Ying Jin$^{2}$ and Xue-Qian Li$^{3,4}$}
\vspace{3.5ex}
\centerline{\small $^{1}$Department of Physics and Mathematical Physics,}
\centerline{\small and Special Research Center for the Subatomic Structure of
Matter,}
\centerline{\small University of Adelaide, SA 5005, Australia}
\centerline{\small $^{2}$Institute of High Energy Physics, Academia Sinica,
        Beijing 100039, P. R. China}
\centerline{\small $^{3}$CCAST(World Laboratory) P.O.Box 8730,
        Beijing 100080, P. R. China}
\centerline{\small $^{4}$Department of Physics, Nankai University, 
Tianjin 300071, P.R. China}

\begin{abstract}
In the heavy quark limit a heavy  baryon which contains two heavy quarks
is believed to be composed of a heavy diquark and a light quark. Based on this 
picture, we evaluate the weak semileptonic decay rates of such baryons. 
The transition form factors between two heavy baryons are associated
with those between two heavy mesons by applying the
superflavor symmetry. The effective vertices
of the W-boson and two heavy diquarks are obtained in
terms of  the Bethe-Salpeter equation. Numerical predictions on these 
semileptonic decay widths are presented and they
will be tested in the future experiments.
\end{abstract}
\vspace{0.5cm}
\end{titlepage}

\vspace{1.0cm}
\section{Introduction}
\hspace{0.29cm} The heavy flavor physics has been an interesting subject for
many years. The meson case has been studied much more intensively both in 
experiments and in theory than the baryon case.
The existence of three valence
quarks in a baryon makes the theoretical study much more
complicated. 
Recently more and more data for heavy baryons which contain one heavy quark
have been accumulated \cite{particle} and in the near
future we may expect even more data from LEP and other experimental groups.
Although we do not have any data for the heavy baryons
containing two heavy quarks (later we will call such baryons $B_{QQ'}$ 
where Q and Q' could be b-quark or c-quark) at present, 
it would be interesting to make predictions
on their properties which will be tested in the future experiments.
In our previous paper \cite{guo} we have 
studied the production of a pair of 
$B_{QQ'}$ in electron-position collisions. It is the aim of the present work
to study the weak semileptonic decays of $B_{QQ'}$. 

        The basic problem is how to deal with the transition form factors
between $B_{QQ'}$ and $B_{QQ''}$ (or $B_{Q''Q'}$) where one flavor transits
(explicitly $b\rightarrow c$) with another heavy quark and the light flavor
remaining unchanged. Since it is determined by the non-perturbative QCD
effects, the solution is by no means trivial.
The heavy quark effective theory (HQET) provides a way to appropriately
simplify the evaluation of the hadronic matrix elements \cite{isgur}
because by applying the HQET we are able to find relations
among the form factors, and consequently reduce the independent number
of these form factors. It is well known that in the heavy quark limit the 
extra symmetries $SU(2)_f\times SU(2)_s$ manifest and the non-perturbative 
effects are attributed to the well-defined Isgur-Wise function 
$\xi(v\cdot v')$, where
$v$ and $v'$ are the four-velocities of the concerned heavy quarks. 

        It is pointed out in our previous paper \cite{guo} that in a
heavy baryon which contains two heavy quarks, these two heavy quarks 
constitute a relatively stable heavy diquark (which will be called $\chi_{QQ'}$
later). This allegation has also been suggested by other authors \cite{Savage}.
The leftover light quark 
moves in the color field induced by the heavy diquark. The size of the
heavy diquark is much smaller compared with the QCD scale $\Lambda_{QCD}$.
In this scenario, the three-body problem is simplified into a two-body
problem.
The ground state heavy diquark can be a spin-1 or spin-0 object. Due to 
the Pauli principle, when Q=Q', the cc- or bb-diquark can 
only be in the spin-1 state while for bc-diquark its spin may be either 0 or 1.
Therefore, from QQ-diquark we can construct a heavy baryon either with spin-$
\frac{3}{2}$ ($B^*_{(QQ)_1}$) or with spin-$\frac{1}{2}$ ($B_{(QQ)_1}$). On the
other hand, from bc-diquark we may have spin-$\frac{1}{2}$ baryon which
is constructed from $\chi_{(bc)_0}$ ($B_{(bc)_0}$) or from $\chi_{(bc)_1}$ 
($B_{(bc)_1}$), and also spin-$\frac{3}{2}$ baryon from
$\chi_{(bc)_1}$ ($B^*_{(bc)_1}$). In the present paper we will study the
weak transition hadronic matrix elements between these heavy baryons and
then give the predictions for the semileptonic decay widths of $B_{QQ'}$. 

Due to the analogue of a heavy meson and a heavy baryon with a heavy diquark,
the superflavor symmetry is applicable to associate the transition matrix
elements between two heavy baryons $B_{QQ'}$ with those between two heavy 
mesons. The superflavor symmetry was first established by 
Georgi and Wise \cite{geor3} for
interchanging a heavy quark and a heavy scalar object, later Carone
\cite{carone} generalized it to the symmetry of interchanging a heavy quark and
a heavy axial vector object. Since the heavy diquark is not really 
point-like with respect to weak transitions, we need to derive the
explicit expressions of the effective vertices
$\chi\chi'W^{\pm}$ by taking into account the inner structure of heavy
diquarks.
Obviously these vertices are associated with the bound state properties of 
heavy diquarks $\chi$ and $\chi'$. Therefore, 
some non-perturbative model has to be adopted.
As in our previous work \cite{guo} we will apply the Bethe-Salpeter (B-S) 
equation model to obtain such vertices.

The paper is organized as the following: In sect.2 we give a detailed 
derivation
of the transition form factors between two heavy diquarks with
a virtual $W-$boson being emitted. Consequently we obtain the effective
currents for heavy diquark weak transitions. 
Then in section 3. we apply superflavor symmetry to give
the formulation for the weak matrix elements between heavy baryons
and the semileptonic decay widths. The numerical results will be presented in
section 4. Finally the last section is devoted to summary and
discussions.

\vspace{1.0cm}
\section{Derivation of the heavy diquark transition form factors }

Since in the heavy quark limit the two heavy quarks in a baryon
constitute a heavy diquark, in the decay process this diquark may be treated
as a color-triple quasi-particle. 
It is noted that for applying the HQET to associate a baryon case to a meson
case, the diquark should be of a point-like
structure, the reason is that all non-perturbative effects are attributed
into a well-defined Isgur-Wise function, therefore the necessary condition
is that the diquark is seen by the light quark as
a point-like color source. However, it
by no means demands that in the weak transition the weak current see a 
point-like structureless object, by contraries, there is
complicated structure
due to the bound state effects of the diquark.
The structure effects
of the heavy diquark should be described by the bound state 
equation. Hence we have to adopt a plausible method to deal with
the diquark structure effects which are governed by the non-perturbative QCD.
In this section we solve the B-S equation \cite{bethe} to obtain the
bound state
wave function of the heavy diquark and then give the transition form factors
between such heavy diquarks in the weak decay processes.

Since the bound state B-S wave functions and the transition form factors 
between
two heavy diquarks are obtained in the same framework, in our formulation 
one does not need to invoke some phenomenological
inputs except the commonly accepted parameters such as $\alpha_s$
and $\kappa$ in the Cornell potential model.

The B-S equation for a heavy diquark can be written in the
following form
\begin{equation}
\label{ji1}
\chi_P(p)=S_1(\lambda_1P+p)\int G(P,p,q)\chi_P(q)
{d^4q\over (2\pi)^4}S_2(\lambda_2P-p),
\end{equation}
where $S_j(j=1,2)$ are the propagators of heavy quark 1 and quark 2
in the diquark
respectively and $G(P,p,q)$ is the B-S equation kernel defined as the
the sum of all the irreducible diagrams concerning the interaction
between the two quarks of the diquark, $\lambda_1={m_1\over
m_1+m_2}$, $\lambda_2={m_2\over m_1+m_2}$, and $m_1$, $m_2$ are the quark 
masses. $P$ is the total momentum of the diquark and can be expressed as 
$P=Mv$ where $M$ is the mass of the diquark and $v$ is its four-velocity.

Using the relation
\begin{equation}
\label{ji2}
S_j(p)=i[{\Lambda^+_j(p_t)\over p_l-W_j+i\epsilon}+
{\Lambda^-_j(p_t)\over p_l+W_j-
i\epsilon}]\rlap/v \;\;\;\;\; (j=1,2)
\end{equation}
where $p_l=p\cdot v$, $p_t=p-p_lv$, $W_j=\sqrt {|p_t|^2+m_j^2}$ and
$\Lambda^{\pm}_j(p_t)={W_j\pm\rlap/v (-p_t+m_j)\over 2W_j}$, Eq.(\ref{ji1})
can be expressed explicitly as
\begin{eqnarray}
\chi^{++}_P(p) && = {-\Lambda_1^+(p_t)\rlap/v\over \lambda_1M+p_l-W_1
+i\epsilon}
\int G(P,p,q)[\chi^{++}(q)+\chi^{--}(q)]{d^4q\over (2\pi)^4}  \nonumber \\
&& {\rlap/v\Lambda^+_2(-p_t)\over p_l+W_2-\lambda_2M-i\epsilon}, \\
\chi^{--}_P(p) && = {-\Lambda_1^-(p_t)\rlap/v\over \lambda_1M+p_l+W_1-i\epsilon}
\int G(P,p,q)[\chi^{++}(q)+\chi^{--}(q)]{d^4q\over (2\pi)^4} \nonumber \\
&& {\rlap/v\Lambda^-_2(-p_t)\over p_l-W_2-\lambda_2M+i\epsilon},
\end{eqnarray}
where $\chi^{\pm\pm}_P(p)=\Lambda^{\pm}_1(p_t)\chi_P(p)\Lambda^{\pm}_2(-p_t)$.

In the heavy quark limit it can be shown that  $\Lambda_1^+(p_t)\approx 
{1+\rlap/v \over 2}$, $\Lambda_2^+(-p_t)
\approx {1+\rlap/v\over 2}$, and $\chi_P^{--}$ is small and negligible.
In the following we will only consider the large component $\chi_P^{++}$.

So for a  scalar or an axial vector diquark, the B-S wave function can 
be written in the forms
$$\chi^S_P(p)={1+\rlap/v\over 2}\sqrt{2M}\phi(p),\;\;\;\;\;\;
\chi^A_P(p)={1+\rlap/v\over 2}\sqrt{2M}\gamma_5\rlap/{\eta}\phi(p).$$
The superscript S and A denote the scalar and axial vector diquark 
respectively and $\eta$ is the polarization vector of the vector diquark.

Now we assume the kernel $G$ to have the form\cite{dai}\cite{guo}
\begin{equation}
\label{kernel}
-iG=1\otimes 1 V_1+\rlap/v\otimes\rlap/v V_2, 
\end{equation}
and
$$V_1(p,q)={8\pi \beta_1\kappa\over [(p_t-q_t)^2+\mu^2]^2}-
(2\pi)^3\delta^3(p_t-q_t) \displaystyle\int{\frac{8\pi\beta_1\kappa}
{(k^2+\mu^2)^2}\frac{d^3k}{(2\pi)^3}}, $$
and 
$$V_2(p,q)=-\displaystyle{\frac{16\pi\beta_2\alpha_s}{3(|p_t-q_t|^2+\mu^2)}},$$
where $V_1$ and $V_2$ are the parts of the kernel associated with the 
scalar confinement and one-gluon-exchange diagram respectively\cite{dai}.
The parameters $\beta_1$ and $\beta_2$ are different for 
various color states. For mesons, $\beta_1=1$, $\beta_2=1$, while for 
color-triplet diquarks,  
$\beta_2$ is directly associated to the color factor caused by the single-gluon
exchange, so should be 0.5. In contrast, $\beta_1$ which is related to
the linear confinement cannot be determined so far and 
we just take it as a free parameter within a range of
$0\sim 1$ in numerical evaluations.
As a matter of fact, later
we pick up two typical values 0.5 and 1 for $\beta_1$ for demonstrating the
influence of the color factor. In fact, the final results are not 
sensitive to its value, so that our predictions made with the value within
a certain range can give rise to a reasonable order of magnitude, even
not a precise number.
The parameters $\kappa$ and $\alpha_s$ are
well determined by fitting  experimental data of heavy meson spectra. 
From the heavy meson
experimental data,  $\kappa=0.18$, $\alpha_s=0.4$\cite{eich}.
After substituting the form of the kernel Eq. (\ref{kernel}) into
Eq. (\ref{ji1}) we have the 
following form of the B-S equation 
\begin{equation}
\label{phi}
{\tilde \phi}(p_t)=\displaystyle{\frac{-1}{M-W_1-W_2}\int(V_1-V_2)
{\tilde \phi}(q_t)\frac{d^3q_t}{(2\pi)^3}}, 
\end{equation}
where  
${\tilde\phi}(p_t)=\int \phi(p)\frac{dp_l}{2\pi}$.
The above equation can be solved out numerically and by applying the relation
between $\phi(p_l, p_t)$ and ${\tilde \phi}(p_t)$ we finally obtain the 
numerical solution of the B-S equation. This solution will be applied to 
calculate the weak transition matrix elements of heavy diquarks.

The weak transition form factors of heavy diquarks are closely associated with
their inner structure.
Namely, to evaluate a transition $b
\rightarrow c$ which are constituent quarks of the initial and final diquarks,
some $Q^2-$dependent form factors would naturally emerge.

The form factors are process-dependent. 
For the semileptonic decay $\chi_{bQ'}(v) \rightarrow \chi_{cQ'} (v')
+l+\bar\nu$ with the light quark being a spectator,
the fundamental vertex $J_{\mu}$ corresponds to a radiation of a
virtual W$-$boson, so that
\begin{equation}
J^{\mu}={g_w\over 2\sqrt 2}V_{cb}^*\bar c\gamma^{\mu}
(1-\gamma_5)b,
\end{equation}
where $g_w$ is the weak coupling constant.

The effective currents $L_{\mu}$ in the expressions of the heavy diquark 
transition matrix elements are calculated by means of the B-S equation
for heavy diquarks.

For scalar or axial-vector diquark transitions, one has the following
four types:
\begin{eqnarray}
\label{ji51}
<M'^S(v')|J_{\mu}|M^S(v)>&=&2\sqrt{MM'}[f_1(v\cdot 
v')v'_\mu+f_2(v\cdot v')v_\mu], \\
\label{ji52}
 <M'^A(v',\eta ')|J^\mu|M^A(v,\eta)>&=&
2\sqrt{MM'}[f_3(v\cdot v')\eta '\cdot\eta v'_\mu+ f_4(v\cdot v')\eta 
'\cdot\eta v_\mu \nonumber \\
&&+f_5(v\cdot v')\eta\cdot v'\eta'\cdot v v'_\mu +f_6(v\cdot v')
\eta\cdot v'\eta '\cdot v v_\mu \nonumber \\
&&+ f_7(v\cdot v')\eta\cdot v' \eta_\mu'+f_8(v\cdot v')\eta '\cdot 
v\eta_\mu \nonumber\\
&&+f_9(v\cdot v')i\epsilon_{\mu\nu\rho\sigma}{\eta'}^\nu\eta^\rho 
{v'}^\sigma \nonumber\\
&&+f_{10}(v\cdot v')i\epsilon_{\mu\nu\rho\sigma}{\eta'}^\nu\eta^\rho 
v^\sigma],\\
\label{ji53}
 <M'^A(\eta' ,v')|J_{\mu}|M^S(v)>&=&2\sqrt{MM'}[f_{11}
\eta'_\mu +f_{12}\eta'\cdot v v'_\mu+f_{13}\eta'\cdot v v_\mu + \nonumber\\
&&f_{14}i\epsilon_{\mu\nu\rho\sigma}{\eta'}^\nu{v'}^\rho v^\sigma],\\
\label{ji54}
<M'^S(v')|J_\mu|M^A(\eta,v)>&=& 2\sqrt{MM'}[f_{15}
\eta_\mu +f_{16}\eta\cdot v' v'_\mu +f_{17}\eta\cdot v' v_\mu + \nonumber\\
&&f_{18}i\epsilon_{\mu\nu\rho\sigma}{\eta}^\nu{v}^\rho {v'}^\sigma].
\end{eqnarray}

On the other hand, the effective matrix elements of heavy diquark transitions
can be expressed by the B-S wave functions in the following 
\begin{equation}
\label{matrix}
<M'(v')|J_{\mu}|M(v)>=\int Tr[{{\bar\chi}^{M'}}_{P'}(p')\Gamma
\chi^{M}_P(p)S^{-1}(p_2)(2\pi)^4\delta^4(p_2-p'_2)
\displaystyle{\frac{d^4p}{(2\pi)^4}\frac{d^4p'}{(2\pi)^4}}]
\end{equation}
where $S(p_2)$ is the propagator of $m_2$ quark and $\Gamma$ is the vertex
of $J_{\mu}$.
$p'_i$ and  $p_i$(i=1,2)  are


\begin{equation}
\begin{array}{ll}
p'_1=\lambda'_1 M'v'+p',~~~~~~~~&p'_2=\lambda'_2 M'v'-p',\\
p_1=\lambda_1 Mv+p ,~~~&p_2=-\lambda_2 Mv-p,\\
\end{array}
\end{equation}
where $M, M'$ are the masses of initial and final diquarks.

So the form factors $f_i$(i=1,...,18) in Eqs.(\ref{ji51}) to (\ref{ji54})
can be expressed as an integral of
the two diquarks' wave functions along with specific coefficients.
The numerical values for the coefficients
$f_i$ (1,...,18) in Eqs.(\ref{ji51}) through (\ref{ji54})
are derived by the combination of Eq. (\ref{matrix}) and Eqs.(\ref{ji51}) 
to (\ref{ji54}). The effective currents $L_\mu$ inducing weak transitions
between heavy diquarks can be expressed as 

\begin{equation}
\label{effective}
L_{\mu}=\sum_{i=1}^{18}f_{i}J_{\lambda}^{(i)}
\end{equation}
where the explicit expressions for $J_{\lambda}^{(i)}$
are given in Eq. (\ref{current}) in Appendix A. For instance, the
terms in $L_{\mu}$ which contribute to
$B_{(bb)_1}\rightarrow B_{(bc)_1}$
are $\sum_{i=3}^{10}f_{i}J_{\lambda}^{(i)}$. 
In next section we will apply the effective currents to calculate the 
hadronic transition matrix elements with the aid of superflavor symmetry.

From the heavy meson experimental data,  $\kappa=0.18$ GeV$^2$,
$\alpha_s=0.4$, $m_b=4.8$ GeV, $m_c=1.45$ GeV.

From the B-S equation, the numerical results of  $M$ ( the heavy diquark mass)
corresponding to the various 
quarks $m_i(i=1,2)$ and $\beta_1$ are listed in Table 1.

\centerline{\bf Table 1 Values of heavy diquark masses}

\begin{center}
\begin{tabular}{|r|r|r|r|r|r|r|}
\hline
$\beta_1$ &0.5 &1 &0.5& 1 &0.5& 1 \\
\hline
$m_1(GeV)$ & 4.8 & 4.8 & 4.8 & 4.8 &1.45 &1.45\\
\hline
$m_2(GeV)$ & 4.8 & 4.8 & 1.45 & 1.45 & 1.45 & 1.45 \\
\hline
$M(GeV)$ &9.68 & 9.74 & 6.46& 6.58& 3.27 & 3.33\\
\hline
\end{tabular}

\end{center}

\vspace{1cm}
\section{ Formulation for the transition matrix elements and decay widths}
\vspace{0.2cm}

\noindent (i) The transition amplitudes

For semileptonic decays, the process can be described as a transition
of a heavy baryon into another heavy baryon radiating a virtual W-boson which
turns into a lepton pair $l\bar\nu$ ($\bar l\nu$). In the process, the
factorization is perfect, so that the total transition amplitude can be
written as
\begin{equation}
T\approx <B'|J_{\alpha}|B>l^{\alpha}({i\over M_W^2}),
\end{equation}
where the contribution from the leptonic current is
$$l^{\alpha}\equiv {g_w\over 2\sqrt 2}\bar u_l(p_l)\gamma^{\alpha}(1-
\gamma_5)v_{(\nu)}(p_{\nu}),\;\;\;\;\;\;\;{\rm for}\;\;\; b\rightarrow cl
\bar\nu_l.$$
At the concerned decay energy scale,
the W-boson  propagator ${i\over q^2-M_W^2}(-g_{\mu\nu}+
q_{\mu}q_{\nu}/M_W^2)$ can be approximated as $-ig_{\mu\nu}/M_W^2$.
In our calculations,
we neglect the lepton masses, because $\tau-$lepton production is hard
to measure, we only discuss the cases of $e^-\bar\nu_e$ and $\mu^-\bar
\nu_{\mu}$ radiation.

Thus we need to derive the forms of the hadronic matrix elements 
$<B'|J_{\mu}|B>$.
The effective currents $L_{\mu}$ are derived in Section 2, so we obtain
the hadronic transition matrix elements $<B'|J_{\mu}|B>$
by calculating $<B'|L_{\mu}|B>$ in the diquark-quark picture.
The scalar or axial-vector diquark is treated as a point-like object 
of color-$\bar 3$  and spin-0 or -1 with definite form factors which 
are reflected in the coefficients $f_i$'s
of the effective currents $L_{\mu}$, and combines with the light quark to
constitute a baryon of spin-1/2 or -3/2. Thus we can use the superflavor
symmetry to evaluate the transition matrix elements at the hadron level.
In this scenario, there is
only one uncertain function which is determined by non-perturbative QCD,
i.e. the Isgur$-$Wise function $\xi(v\cdot v')$, unlike the case for
transitions between light baryons where there are many form
factors. Therefore, here we may expect to reduce the uncertainty and improve
the prediction power, and it exactly is the advantage of employing the
superflavor symmetry.

In the scenario of superflavor symmetry \cite{geor3}, the wave function for
a baryon consisting of a scalar diquark would be
\begin{equation}
\tilde{\Psi}_X=\left( \begin{array}{c} 0\\
u^TC/\sqrt{2M_X}   
\end{array} \right),
\end{equation}
where $M_X$ is the mass of the scalar diquark and $C$ is the charge-conjugation
operator satisfying $C^{-1}\gamma_{\mu}^TC=-\gamma_{\mu}$. For the spin-1
diquark case \cite{carone},
\begin{equation}
\tilde\Psi_{1/2}(v)={1\over\sqrt{6M_A}}\left( \begin{array}{c}
0\\
u^TC\sigma^{\mu\beta}v_{\beta}\gamma_5
\end{array} \right),
\end{equation}
and
\begin{equation}
\tilde\Psi_{3/2}={1\over\sqrt 2M_A}\left( \begin{array}{c}
0\\ \psi^{\mu T}C
\end{array} \right),
\end{equation}
where $M_A$ is the mass of the spin-1 diquark and the subscripts 1/2 and 3/2
denote the spins of baryons.

Thus the hadronic transition matrix element can be obtained as
\begin{equation}
T_{\mu}\equiv
<B'_{J'}(v')|L_{\mu}|B_J(v)>=-\xi(v\cdot v')Tr[\overline{\tilde\Psi}'_{J'}
(v')\sum_{i}f_i\Gamma_i\tilde{\Psi}_J(v)],
\end{equation}
where $\Gamma_i$'s are the corresponding vertices in the effective current 
$L_{\mu}$. 
In Ref.\cite{geor3}, the authors presented some transition matrix elements
with certain effective currents, instead, here our effective currents
correspond to the weak interaction.\footnote{For evaluating a radiative decay,
one can have similar effective currents with only small changes from
that given for weak interactions.} The explicit expressions for various
hadronic transition matrix elements $T_{i\mu} (i=1,...12)$
are listed in Eqs.(\ref{t1}) to (\ref{t12}) in Appendix A.\\

\noindent (ii) The amplitude square

To calculate the cross section, we need to take square of the amplitudes
which are given in (i) and Appendix A. In the derivations we use the following
relations 
\begin{equation}
\label{dog}
\sum_{s}u(v,s)\bar u(v,s)={\rlap /v+1\over 2},
\end{equation}
\begin{equation}
\label{cat}
\sum_{s}\Psi_{\lambda}(v,s)\bar\Psi_{\delta}(v,s)=
{\rlap /v+1\over 2}[-g_{\lambda\delta}
+{1\over 3}\gamma_{\lambda}\gamma_{\delta}+{1\over 3}
(\gamma_{\lambda}v_{\delta}
-\gamma_{\delta}v_{\lambda})+{2\over 3}v_{\lambda}v_{\delta}],
\end{equation}
for heavy baryons of spin 1/2 and 3/2 respectively and we neglect the lepton
masses for simplicity.
\begin{eqnarray}
\sum_{s}u_{(l)}(p_3,s)\bar u_{(l)}(p_3,s) &=& (\rlap /p_3+m_l)
\approx \rlap /p_3,\\
\sum_{s}v_{(\nu)}(p_4,s)\bar v_{(\nu)}(p_4,s) &=& \rlap /p_4.
\end{eqnarray}
It is noted that here we adopt the above conventions for the
heavy baryons and leptons, because at the limit $m_l\sim 0$ this choice
provides us with much convenience (also see below for the integration
over the final state phase space).

Then the amplitude square $|T(B_J(v)\rightarrow B_{J'}'(v')+l\bar{\nu})|^2$
can be obtained. The results are given in Appendix B.\\

\noindent (iii) The integration over the final state phase space

To obtain the partial decay width, one needs to integrate out the phase 
space of the
three-body final state. In the limit of $m_l\sim 0$, the integration becomes
much simplified.

It is easy to notice that the amplitude square can be written in a
general form
\begin{eqnarray}
 \sum_{spins}|T_{i\lambda}l^{\lambda}|^2{1\over M_W^4}
&\equiv &
F_1(p_1\cdot p_2)(p_3\cdot p_4)+F_2(p_1\cdot p_3)(p_2\cdot p_4)+
F_3(p_3\cdot p_1)(p_4\cdot p_1) \nonumber \\
&+& F_4(p_3\cdot p_2)(p_4\cdot p_2)+F_5(p_2\cdot p_3)(p_1\cdot p_4)
+F_6(p_3\cdot p_4),
\end{eqnarray}
where $p_3$ and $p_4$ are the momenta of emitted lepton and neutrino, 
$p_1=mv$ is the decaying baryon momentum, so can be $m(1, \vec 0)$ and
$p_2=m'v'$ is the momentum of the decay product which should be integrated
over, $m$ and $m'$ are the masses of initial and final baryons respectively,
and $T_{i\lambda}$ ($i$=1,...,18) are given in Apendix A.

Accordingly, the integration of the final state phase space should be
properly written according to the above conventions. We write the expressions
down explicitly,

\begin{eqnarray}
\Gamma &=& {1\over (2s+1)}\int {d^3p_2\over (2\pi)^3}{m_2\over E_2}\int{d^3p_3
\over (2\pi)^3}{1\over 2E_3}\int{d^3p_4\over (2\pi)^3}{1\over 2E_4}
\sum_{spins}|T_{i\lambda}l^{\lambda}|^2\frac{1}{M_{W}^{4}} \nonumber \\
&& \cdot (2\pi)^4\delta^4(p_1-p_2-p_3-p_4)
\end{eqnarray}
where all $p_i'$s are defined above.

Thus after a simple manipulation,
the final form of the decay width can be written (in the following expression
the spin factor $2s+1=2$ for the spin-1/2 baryon decay while for spin-3/2 
baryon decay $2s+1=4$) as
\begin{eqnarray}
\label{baryon}
\Gamma &=& {m'\over 16(2s+1)\pi^3}\int_0^{(m-m')^2} ds_2\{ {F_1\over 16 m^2}
s_2[m^2+m'^2-s_2] \nonumber \\
&& +{F_2+F_5\over 96 m^2}[(m^2-m'^2+s_2)(m^2-m'^2-s_2)+s_2(m^2+m'^2-s_2)]
\nonumber \\
&& +{F_3\over 48}[{(m^2+s_2-m'^2)^2\over 2m^2}+s_2]
+{F_4\over 48 m^2}[(m^2-m'^2-s_2)^2+m'^2s_2] \nonumber \\
&& +{F_6\over 8m^2} s_2 \}\cdot\lambda^{1/2}(m^2, m'^2, s_2)
\end{eqnarray}
where 
$$\lambda (a,b,c)\equiv a^2+b^2+c^2-2ab-2bc-2ca, $$
and $F_1$ through $F_6$ are given in the expressions of $\sum
|T_{i\alpha} l^{\alpha}|^2/M_W^4$
by rearranging the corresponding terms so that they are expressed in terms of
the form factors $f_i (i=1,..., 18)$ which have been calculated in the B-S
approach. The concrete forms are
obtained by running the REDUCE
computer programs and it is a very lengthy and tedious
procedure. The relations between $F_i (i=1,...,6)$ and $f_i (i=1,...,18)$
are very complicated and we will not list them here.

\vspace{1cm}
\section{Numerical results}

In the relations between $F_i (i=1,...,6)$ and $f_i (i=1,...,18)$ there is an
uncertain function, the Isgur-Wise function. Its behavior is controlled 
by non-perturbative QCD effects which have to be dealt with in some
phenomenological model. Because in the heavy quark limit, the spin of the
heavy quark has no effects on the dynamics inside the hadron one expects that
the Isgur-Wise function is totally determined by the light degrees of freedom.
Therefore, the Isgur-Wise function between the transition 
of the heavy baryons consisting of two heavy quarks
should be the same as that of the corresponding heavy mesons. 
Actually this is the plausibility
of applying the superflavor symmetry. Hence we can simply use the form
of the Isgur-Wise function for $B \rightarrow D$ in our numerical calculations
for the decay width of heavy baryons which contain two heavy quarks. There
are some model calculations for the Isgur-Wise function for $B \rightarrow D$
\cite{rosner}\cite{neubert}\cite{abd}. In our following numerical calculations
we will use the following simple form given in \cite{rosner}
 
\begin{equation}
\label{ros}
\xi (\omega)=\frac{1}{1-\omega^2/\omega_{0}^{2}},
\end{equation}
where the constant is taken to be $\omega_0 =1.24$. It is noted that different
forms of the Isgur-Wise function will give somehow
different predictions. However our numerical computations show that with
various Isgur-Wise function forms
the order is not changed. The numerical results for the
semileptonic decay widths for different processes are listed in Table 2.
From the numerical results in Table 2 we can see that the decay widths
are around the order $10^{-13} \sim 10^{-14}$s$^{-1}$. 
It can also be seen that the results are
insensitive to the parameter $\beta_1$.

\newpage
\centerline{\bf Table 2 Semileptonic decay widths of $B_{QQ'}$(s$^{-1}$)}

\begin{center}
\begin{tabular}{|c|c|c|}
\hline
$\beta_1$ &1 &0.5 \\
\hline
$B_{(bb)_1}\rightarrow B_{(bc)_1}$ &$2.85\times 10^{-13}$ &$2.78\times
10^{-13}$ \\
\hline
$B_{(bb)_1}\rightarrow B_{(bc)_0}$&$4.28\times 10^{-14}$ &$4.81\times
10^{-14}$ \\
\hline
$B_{(bb)_1}\rightarrow B^*_{(bc)_1}$&$2.72\times 10^{-13}$&$2.69\times
10^{-13}$\\
\hline
$B^*_{(bb)_1}\rightarrow B^*_{(bc)_1}$&$1.29\times 10^{-13}$&$1.41\times
10^{-13}$\\
\hline
$B^*_{(bb)_1}\rightarrow B_{(bc)_1}$&$5.20\times 10^{-13}$&$5.15\times
10^{-13}$\\
\hline
$B^*_{(bb)_1}\rightarrow B_{(bc)_0}$&$8.57\times 10^{-14}$&$9.61\times 
10^{-14}$\\
\hline
$B^*_{(bc)_1}\rightarrow B^*_{(cc)_1}$&$1.72\times
10^{-13}$&$1.68\times 10^{-13}$\\
\hline
$B^*_{(bc)_1}\rightarrow B_{(cc)_1}$&$2.75\times 10^{-13}$&$2.81\times
10^{-13}$\\
\hline
$B_{(bc)_1}\rightarrow B^*_{(cc)_1}$&$1.41\times 10^{-13}$&$1.43\times 
10^{-13}$\\
\hline
$B_{(bc)_1}\rightarrow B_{(cc)_1}$&$8.93\times 10^{-14}$&$9.32\times
10^{-14}$\\
\hline
$B_{(bc)_0}\rightarrow B^*_{(bc)_1}$&$2.88\times 10^{-13}$&$2.91\times 
10^{-13}$\\
\hline
$B_{(bc)_0}\rightarrow B_{(cc)_1}$&$7.76\times 10^{-14}$&$7.82\times
10^{-14}$\\
\hline
\end{tabular}

\end{center}

\vspace{1cm}
\section{Summary and discussions}

	In the present work we discuss the weak transitions
between heavy baryons which consist of two heavy quarks in the heavy quark
limit. The three-body system is simplified into a two-body system of a heavy
diquark and a light quark. In the heavy quark limit,
the heavy diquark is a point-like spin-0 or spin-1 object
and the light quark is blind to the spin and flavor of the heavy diquark.
With the help of the superflavor symmetry the matrix elements between 
heavy baryons and those between
heavy mesons are related to each other and they can be described by the
same Isgur-Wise function.
To deal with the weak transitions between heavy
diquarks we work in the B-S equation approach. We obtain the numerical 
solutions of the B-S equation by assuming the kernel which contains
linear scalar confinement and one-gluon-exchange vector terms. The numerical
solutions are used to obtain the effective currents between two heavy diquarks.
These effective currents are expressed in terms of the coefficients
$f_i (i=1,...,18)$ which can be solved out numerically from the B-S equation
for heavy diquarks.
Then the weak transition matrix elements between heavy baryons are expressed
in terms of the Isgur-Wise function and $f_i (i=1,...,18)$. Consequently
we give the predictions for the semileptonic decay widths for all the
possible twelve decay channels between two heavy baryons.
The decay widths are around the order $10^{-13} \sim 10^{-14}$s$^{-1}$. These
predictions will be tested in the future experiments.

There are some uncertainties in our work. First, as we have been
working in the
heavy quark limit, all $1/m_Q$ corrections are ignored.
In the heavy quark limit
physics is greatly simplified and we have only one unknown function, the 
Isgur-Wise function. Therefore, if one wishes to make a precise comparison of
the theoretically calculated numbers with data, the
$1/m_Q$ especially $1/m_c$ corrections must be taken into account. 
Besides this approximation when we calculate the
effective currents between two heavy diquarks we work in the B-S equation 
approach in which the most uncertain point is the kernel which depends on
non-perturbative QCD effects. Motivated by potential model we use the
simple form which has linear scalar confinement and one-gluon-exchange terms.
In the confinement part the parameter $\beta_1$ is not fixed  and we pick up 
two typical values as 0.5 and 1. Fortunately, the decay widths are 
insensitive to
this parameter.
The Isgur-Wise function is another uncertain
point since it is also controlled by non-perturbative QCD dynamics 
between heavy diquark and light quark and thus
its evaluation is model dependent. To get the numerical results we use the
simple form
obtained in Ref. \cite{rosner}. Different forms for the Isgur-Wise function
may result in different decay widths. However, the order is not changed.

\vspace{1cm}

\noindent {\bf Acknowledgment}:
\vspace{2mm}

This work was supported in part by the Australian Research Council and
the National Science Foundation of China.

\newpage

\noindent {\bf Appendix A. The hadronic transition matrix elements}
\vspace{0.3cm}

\noindent (1) $B_{(bb)_1}\rightarrow B_{(bc)_1}$
\begin{eqnarray}
\label{t1}
T_{1\lambda} &=& <B^A_{(bc)}({1\over 2})|L_{\lambda}|B^A_{(bb)}({1\over 2})>
 \nonumber\\
 &=& 2\sqrt{M_1M_2}<B^A_{(bc)}({1\over 2})|f_3J_{\lambda}^{(3)}
+f_4J_{\lambda}^{(4)}+f_5J_{\lambda}^{(5)}+f_6J_{\lambda}^{(6)}+ \nonumber\\
&& f_7J_{\lambda}^{(7)}+f_8J_{\lambda}^{(8)} +f_9J_{\lambda}^{(9)}
+f_{10}J_{\lambda}^{(10)}|B^A_{(bb)}({1\over 2})> \nonumber \\
&=& {1\over 3}\xi(v\cdot v') [(-f_3v'_{\lambda}
-f_4v_{\lambda})(2+v\cdot v') +(f_5v'_{\lambda} 
 +f_6v_{\lambda})(1-(v\cdot v')^2) \nonumber \\
&& -(f_7v'_{\lambda}+f_8v_{\lambda})(1+v\cdot v')]\bar u'u 
+(f_7+f_8)(1+v\cdot v')\bar u'\gamma_{\lambda}u \nonumber \\
&&+ i(f_9v^{\rho}v'^{\sigma}-
f_{10}v'^{\rho}v^{\sigma})
\epsilon_{\lambda\delta\rho\sigma}\bar u'\gamma^{\delta}u \nonumber\\
&& -i(f_9v'^{\sigma}+f_{10}v^{\sigma})\epsilon_{\lambda\delta\rho\sigma}
\bar u'\gamma^{\delta}\gamma^{\rho}u \}.
\end{eqnarray}

\noindent (2) $B_{(bb)_1}\rightarrow B_{(bc)_0}$
\begin{eqnarray}
T_{2\lambda} &=& <B^S_{(bc)}({1\over 2})|L_{\lambda}|B^A_{(bb)}({1\over 2})>
 \nonumber\\
 &=& 2\sqrt{M_1M_2}<B^S_{(bc)}({1\over 2})|f_{15}J_{\lambda}^{(15)}
+f_{16}J_{\lambda}^{(16)}+f_{17}J_{\lambda}^{(17)}
+f_{18}J_{\lambda}^{(18)}|B^A_{(bb)}({1\over 2})> \nonumber \\
&=& {i\over\sqrt 3}\xi(v\cdot v')\{[(1+v\cdot v')(f_{16}v'_{\lambda}
+f_{17}v_{\lambda})+f_{15}v_{\lambda}]\bar u'\gamma_5u \nonumber\\
&& +f_{15}\bar u'\gamma_{\lambda}\gamma_5u-if_{18}\epsilon_{\lambda\delta\rho
\sigma}\bar u'\gamma_5\gamma^{\delta}u \}.
\end{eqnarray}

\noindent (3) $B_{(bb)_1}\rightarrow B^*_{(bc)_1}$
\begin{eqnarray}
T_{3\lambda} &=& <B^A_{(bc)}({3\over 2})|L_{\lambda}|B^A_{(bb)}({1\over 2})>
 \nonumber\\
 &=& 2\sqrt{M_1M_2}<B^A_{(bc)}({3\over 2})|f_3J_{\lambda}^{(3)}
+f_4J_{\lambda}^{(4)}+f_5J_{\lambda}^{(5)}+f_6J_{\lambda}^{(6)}+ \nonumber\\
&& f_7J_{\lambda}^{(7)}+f_8J_{\lambda}^{(8)} +f_9J_{\lambda}^{(9)}
+f_{10}J_{\lambda}^{(10)}|B^A_{ (bb)}({1\over 2})> \nonumber \\
&=& {i\over \sqrt 3}\xi(v\cdot v') \{ [f_3v'_{\lambda}v_{\delta}
+f_4v_{\lambda}v_{\delta}+(f_5v'_{\lambda}v_{\delta} 
 +f_6v_{\lambda}v_{\delta})(1+v\cdot v') \nonumber \\
&& +f_8v_{\lambda}v_{\delta}]\bar \Psi'^{\delta}\gamma_5 u 
\nonumber \\
&& +(f_3v'_{\lambda}+f_4v_{\lambda})\bar\Psi'^{\delta}\gamma_{\delta}\gamma_5 u
+f_8v_{\delta}\bar\Psi'^{\delta}\gamma_{\lambda}\gamma_5u \nonumber \\
&&+ if_9v^{\rho}v'^{\sigma}\epsilon_{\lambda\delta\rho\sigma}
\bar\Psi'^{\delta}\gamma_5u 
 +i(f_9v'^{\sigma}+f_{10}v^{\sigma})\epsilon_{\lambda\delta\rho\sigma}
\bar \Psi'^{\delta}\gamma^{\rho}\gamma_5u \}.
\end{eqnarray}

\noindent (4) $B^*_{(bb)_1}\rightarrow B^*_{(bc)_1}$
\begin{eqnarray}
T_{4\lambda} &=& <B^A_{(bc)}({3\over 2})|L_{\lambda}|B^A_{(bb)}({3\over 2})>
 \nonumber\\
 &=& 2\sqrt{M_1M_2}<B^A_{(bc)}({1\over 2})|f_3J_{\lambda}^{(3)}
+f_4J_{\lambda}^{(4)}+f_5J_{\lambda}^{(5)}+f_6J_{\lambda}^{(6)}+ \nonumber\\
&& f_7J_{\lambda}^{(7)}+f_8J_{\lambda}^{(8)} +f_9J_{\lambda}^{(9)}
+f_{10}J_{\lambda}^{(10)}|B^A_{(bb)}({1\over 2})> \nonumber \\
&=& \xi(v\cdot v') \{ (f_3v'_{\lambda}
+f_4v_{\lambda})\bar\Psi'^{\delta}\Psi_{\delta}+
(f_5v'_{\lambda}v'_{\nu}v_{\mu} 
 +f_6v_{\lambda}v'_{\nu}v_{\mu})\bar\Psi'^{\mu}\psi^{\nu} \nonumber \\
&& +(f_7v'_{\delta}+f_8v_{\delta})\bar\Psi'^{\delta}\Psi_{\lambda}
+ i(f_9v'^{\sigma}+f_{10}v^{\sigma})
\epsilon_{\lambda\delta\rho\sigma}\bar \Psi'^{\delta}\Psi^{\rho} \}.
\end{eqnarray}

\noindent (5) $B^*_{(bb)_1}\rightarrow B_{(bc)_1}$
\begin{eqnarray}
T_{5\lambda} &=& <B^A_{(bc)}({1\over 2})|L_{\lambda}|B^A_{(bb)}({3\over 2})>
 \nonumber\\
 &=& 2\sqrt{M_1M_2}<B^A_{(bc)}({1\over 2})|f_3J_{\lambda}^{(3)}
+f_4J_{\lambda}^{(4)}+f_5J_{\lambda}^{(5)}+f_6J_{\lambda}^{(6)}+ \nonumber\\
&& f_7J_{\lambda}^{(7)}+f_8J_{\lambda}^{(8)} +f_9J_{\lambda}^{(9)}
+f_{10}J_{\lambda}^{(10)}|B^A_{(bb)}({1\over 2})> \nonumber \\
&=& {i\over \sqrt 3}\xi(v\cdot v')\{ [(f_3v'_{\lambda}v'_{\alpha}
+f_4v_{\lambda}v'_{\alpha}) +(f_5v'_{\lambda}v'_{\alpha} 
 +f_6v_{\lambda}v'_{\alpha})(1+v\cdot v')
+f_7v'_{\lambda}v'_{\alpha}]\bar u'\gamma_5\Psi^{\alpha} \nonumber \\
&&+ f_8(1+v\cdot v')\bar u'\gamma_5\Psi_{\lambda}-(f_3v'_{\lambda}+
f_4v_{\lambda})\bar u'\gamma_{\alpha}\gamma_5\Psi^{\alpha}
-f_7v'_{\alpha}\bar u'\gamma_{\lambda}\gamma_5\Psi^{\alpha} \nonumber\\ 
&& +if_{10}v'^{\delta}v^{\sigma}
\epsilon_{\lambda\delta\rho\sigma}\bar u'\gamma_5\Psi^{\rho}
 -i(f_9v'^{\sigma}+f_{10}v^{\sigma})\epsilon_{\lambda\delta\rho\sigma}
\bar u'\gamma^{\delta}\gamma_5\Psi^{\rho} \}.
\end{eqnarray}

\noindent (6)  $B^*_{(bb)_1}\rightarrow B_{(bc)_0}$
\begin{eqnarray}
T_{6\lambda} &=& <B^S_{(bc)}({1\over 2})|L_{\lambda}|B^A_{(bb)}({3\over 2})>
 \nonumber\\
 &=& 2\sqrt{M_1M_2}<B^S_{(bc)}({1\over 2})|f_{15}J_{\lambda}^{(15)}
+f_{16}J_{\lambda}^{(16)}+f_{17}J_{\lambda}^{(17)}
+f_{18}J_{\lambda}^{(18)}|B^A_{(bb)}({3\over 2})> \nonumber \\
&=& \xi(v\cdot v')\{ f_{15}\bar u'\Psi_{\lambda}
+(f_{16}v'_{\delta}v'_{\lambda}+f_{17}v'_{\delta}v_{\lambda})\bar u'
\Psi^{\delta}\nonumber \\
&& +if_{18}\epsilon_{\lambda\delta\rho\sigma}v^{\rho}v'^{\sigma})\bar u'
\Psi^{\delta} \}.
\end{eqnarray}

\noindent (7) $B^*_{(bc)_1}\rightarrow B^*_{(cc)_1}$
\begin{eqnarray}
T_{7\lambda} &=& <B^A_{(cc)}({3\over 2})|L_{\lambda}|B^A_{(bc)}({3\over 2})>
 \nonumber\\
 &=& 2\sqrt{M_1M_2}<B^A_{(bc)}({1\over 2})|f_3J_{\lambda}^{(3)}
+f_4J_{\lambda}^{(4)}+f_5J_{\lambda}^{(5)}+f_6J_{\lambda}^{(6)}+ \nonumber\\
&& f_7J_{\lambda}^{(7)}+f_8J_{\lambda}^{(8)} +f_9J_{\lambda}^{(9)}
+f_{10}J_{\lambda}^{(10)}|B^A_{(bb)}({1\over 2})> \nonumber \\
&=& \xi(v\cdot v') \{ (f_3v'_{\lambda}
+f_4v_{\lambda})\bar\Psi'^{\delta}\Psi_{\delta}+
(f_5v'_{\lambda}v'_{\nu}v_{\mu} 
 +f_6v_{\lambda}v'_{\nu}v_{\mu})\bar\Psi'^{\mu}\psi^{\nu} \nonumber \\
&& +(f_7v'_{\delta}+f_8v_{\delta})\bar\Psi'^{\delta}\Psi_{\lambda}
+ i(f_9v'^{\sigma}+f_{10}v^{\sigma})
\epsilon_{\lambda\delta\rho\sigma}\bar \Psi'^{\delta}\Psi^{\rho} \}.
\end{eqnarray}

\noindent (8) $B^*_{(bc)_1}\rightarrow B_{(cc)_1}$
\begin{eqnarray}
T_{8\lambda} &=& <B^A_{(cc)}({1\over 2})|L_{\lambda}|B^A_{(bc)}({3\over 2})>
 \nonumber\\
 &=& 2\sqrt{M_1M_2}<B^A_{(bc)}({1\over 2})|f_3J_{\lambda}^{(3)}
+f_4J_{\lambda}^{(4)}+f_5J_{\lambda}^{(5)}+f_6J_{\lambda}^{(6)}+ \nonumber\\
&& f_7J_{\lambda}^{(7)}+f_8J_{\lambda}^{(8)} +f_9J_{\lambda}^{(9)}
+f_{10}J_{\lambda}^{(10)}|B^A_{(bb)}({1\over 2})> \nonumber \\
&=& {i\over \sqrt 3}\xi(v\cdot v')\{ [(f_3v'_{\lambda}v'_{\alpha}
+f_4v_{\lambda}v'_{\alpha}) +(f_5v'_{\lambda}v'_{\alpha} 
 +f_6v_{\lambda}v'_{\alpha})(1+v\cdot v')
+f_7v'_{\lambda}v'_{\alpha}]\bar u'\gamma_5\Psi^{\alpha} \nonumber \\
&&+ f_8(1+v\cdot v')\bar u'\gamma_5\Psi_{\lambda}-(f_3v'_{\lambda}+
f_4v_{\lambda})\bar u'\gamma_{\alpha}\gamma_5\Psi^{\alpha}
-f_7v'_{\alpha}\bar u'\gamma_{\lambda}\gamma_5\Psi^{\alpha} \nonumber\\ 
&& +if_{10}v'^{\delta}v^{\sigma}
\epsilon_{\lambda\delta\rho\sigma}\bar u'\gamma_5\Psi^{\rho}
 -i(f_9v'^{\sigma}+f_{10}v^{\sigma})\epsilon_{\lambda\delta\rho\sigma}
\bar u'\gamma^{\delta}\gamma_5\Psi^{\rho} \}.
\end{eqnarray}

\noindent (9) $B_{(bc)_1}\rightarrow B^*_{(cc)_1}$
\begin{eqnarray}
T_{9\lambda} &=& <B^A_{(bc)}({3\over 2})|L_{\lambda}|B^A_{(bb)}({1\over 2})>
 \nonumber\\
 &=& 2\sqrt{M_1M_2}<B^A_{(bc)}({3\over 2})|f_3J_{\lambda}^{(3)}
+f_4J_{\lambda}^{(4)}+f_5J_{\lambda}^{(5)}+f_6J_{\lambda}^{(6)}+ \nonumber\\
&& f_7J_{\lambda}^{(7)}+f_8J_{\lambda}^{(8)} +f_9J_{\lambda}^{(9)}
+f_{10}J_{\lambda}^{(10)}|B^A_{ (bb)}({1\over 2})> \nonumber \\
&=& {i\over \sqrt 3}\xi(v\cdot v') \{ [f_3v'_{\lambda}v_{\delta}
+f_4v_{\lambda}v_{\delta}+(f_5v'_{\lambda}v_{\delta} 
 +f_6v_{\lambda}v_{\delta})(1+v\cdot v') \nonumber \\
&& +f_8v_{\lambda}v_{\delta}]\bar \Psi'^{\delta}\gamma_5 u 
\nonumber \\
&& -(f_3v'_{\lambda}+f_4v_{\lambda})\bar\Psi'^{\delta}\gamma_5\gamma_{\delta} u
-f_8v_{\delta}\bar\Psi'^{\delta}\gamma_5\gamma_{\lambda}u \nonumber \\
&&+ if_9v^{\rho}v'^{\sigma}\epsilon_{\lambda\delta\rho\sigma}
\bar\Psi'^{\delta}\gamma_5u 
 -i(f_9v'^{\sigma}+f_{10}v^{\sigma})\epsilon_{\lambda\delta\rho\sigma}
\bar \Psi'^{\delta}\gamma_5\gamma^{\rho}u \}.
\end{eqnarray}

\noindent (10) $B_{(bc)_1}\rightarrow B_{(cc)_1}$
\begin{eqnarray}
T_{10\lambda} &=& <B^A_{(bc)}({1\over 2})|L_{\lambda}|B^A_{(bb)}({1\over 2})>
 \nonumber\\
 &=& 2\sqrt{M_1M_2}<B^A_{(bc)}({1\over 2})|f_3J_{\lambda}^{(3)}
+f_4J_{\lambda}^{(4)}+f_5J_{\lambda}^{(5)}+f_6J_{\lambda}^{(6)}+ \nonumber\\
&& f_7J_{\lambda}^{(7)}+f_8J_{\lambda}^{(8)} +f_9J_{\lambda}^{(9)}
+f_{10}J_{\lambda}^{(10)}|B^A_{(bb)}({1\over 2})> \nonumber \\
&=& {1\over 3}\xi(v\cdot v') [(-f_3v'_{\lambda}
-f_4v_{\lambda})(2+v\cdot v') +(f_5v'_{\lambda} 
 +f_6v_{\lambda})(1-(v\cdot v')^2) \nonumber \\
&& -(f_7v'_{\lambda}+f_8v_{\lambda})(1+v\cdot v')]\bar u'u 
+(f_7+f_8)(1+v\cdot v')\bar u'\gamma_{\lambda} u \nonumber \\
&&+ i(f_9v^{\rho}v'^{\sigma}-
f_{10}v'^{\rho}v^{\sigma})
\epsilon_{\lambda\delta\rho\sigma}\bar u'\gamma^{\delta}u \nonumber\\
&& -i(f_9v'^{\sigma}+f_{10}v^{\sigma})\epsilon_{\lambda\delta\rho\sigma}
\bar u'\gamma^{\delta}\gamma^{\rho}u \}.
\end{eqnarray}

\noindent (11) $B_{(bc)_0}\rightarrow B^*_{(bc)_1}$
\begin{eqnarray}
T_{11\lambda} &=& <B^A_{(cc)}({3\over 2})|L_{\lambda}|B^S_{(bc)}({1\over 2})>
 \nonumber\\
 &=& 2\sqrt{M_1M_2}<B^A_{(cc)}({3\over 2})|f_{11}J_{\lambda}^{(11)}
+f_{12}J_{\lambda}^{(12)}+f_{13}J_{\lambda}^{(13)}
+f_{14}J_{\lambda}^{(14)}|B^A_{(bc)}({1\over 2})> \nonumber \\
&=& \xi(v\cdot v')\{ f_{11}\bar \Psi'_{\lambda}u
+(f_{12}v_{\delta}v'_{\lambda}+f_{13}v_{\delta}v_{\lambda})\bar 
\Psi'^{\delta} u \nonumber \\
&& +if_{14}\epsilon_{\lambda\delta\rho\sigma}v'^{\rho}v^{\sigma})\bar
\Psi'^{\delta} u \}.
\end{eqnarray}

\noindent (12) $B_{(bc)_0}\rightarrow B_{(cc)_1}$
\begin{eqnarray}
\label{t12}
T_{12\lambda} &=& <B^A_{(cc)}({1\over 2})|L_{\lambda}|B^S_{(bc)}({1\over 2})>
 \nonumber\\
 &=& 2\sqrt{M_1M_2}<B^A_{(cc)}({1\over 2})|f_{11}J_{\lambda}^{(11)}
+f_{12}J_{\lambda}^{(12)}+f_{13}J_{\lambda}^{(13)}
+f_{14}J_{\lambda}^{(14)}|B^A_{(bc)}({1\over 2})> \nonumber \\
&=& {i\over\sqrt 3}\xi(v\cdot v')\{ [f_{11}v'_{\lambda}+(f_{12}
v'_{\lambda}+f_{13}v_{\lambda})(1+v\cdot v')]\bar u'\gamma_5 u \nonumber\\
&& -f_{11}\bar u'\gamma_{\lambda}\gamma_5 u
-if_{14}\epsilon_{\lambda\delta\rho\sigma}v'^{\rho}v^{\sigma}\bar
u'\gamma^{\delta}\gamma_5 u \}.
\end{eqnarray}

In the Eqs.(\ref{t1}) to (\ref{t12}) we define

\begin{equation}
\label{current}
\begin{array}{ll}
J_{\lambda}^{(1)} = \chi_f^{\dagger}(v')  v'_{\lambda}\chi_i(v), &
J_{\lambda}^{(2)} = \chi_f^{\dagger}(v') v_{\lambda}\chi_i(v), \\
J_{\lambda}^{(3)} = {\chi_f^{\mu}}^{\dagger}(v') v'_{\lambda}\chi_{i\mu}(v),&
J_{\lambda}^{(4)} = {\chi_f^{\mu}}^{\dagger}(v') v_{\lambda}\chi_{i\mu}(v), \\
J_{\lambda}^{(5)} = ({\chi_f^{\mu}}^{\dagger}(v')v_{\mu}), 
v'_{\lambda}(v'_{\nu}\chi_i^{\nu}(v)), &
J_{\lambda}^{(6)} = ({\chi_f^{\mu}}^{\dagger}(v')v_{\mu})
v_{\lambda}(v'_{\nu}\chi_i^{\nu}(v)), \\
J_{\lambda}^{(7)} = {\chi_{f\lambda}}^{\dagger}(v')
(v'_{\nu}\chi_i^{\nu}(v)),&
J_{\lambda}^{(8)} = ({\chi_f^{\mu}}^{\dagger}(v')v_{\mu})
\chi_i^{\lambda}(v), \\
 J_{\lambda}^{(9)} = i\epsilon_{\lambda\delta\rho\sigma}
{\chi_f^{*\delta}}^{\dagger}(v')\chi_i(v)v^{\rho}v'^{\sigma}, &
J_{\lambda}^{(10)} = i\epsilon_{\lambda\delta\rho\sigma}
{\chi_f^{*\delta}}^{\dagger}(v')\chi_i(v)v'^{\rho}v^{\sigma}, \\
J_{\lambda}^{(11)} = {\chi_{f\lambda}}^{\dagger}(v')\chi_i(v), &
J_{\lambda}^{(12)} = ({\chi_f^{\mu}}^{\dagger}(v')v_{\mu})v'_{\lambda}
\chi_i(v), \\
J_{\lambda}^{(13)} = ({\chi_f^{\mu}}^{\dagger}(v')v_{\mu})v_{\lambda}
\chi_i(v), &
J_{\lambda}^{(14)} = i\epsilon_{\lambda\delta\rho\sigma}{\chi_f^{\delta}}
^{\dagger}(v')v'^{\rho}v^{\sigma}\chi_i(v), \\
J_{\lambda}^{(15)} = \chi_f^{\dagger}(v')\chi_{i\lambda}(v), &
J_{\lambda}^{(16)} = \chi_f^{\dagger}(v')v'_{\lambda}v'^{\mu}\chi_{i\mu}(v)
 \\
J_{\lambda}^{(17)} = \chi_f^{\dagger}(v')v_{\lambda}v'^{\mu}\chi_{i\mu}(v), &
J_{\lambda}^{(18)} = i\epsilon_{\lambda\delta\rho\sigma}\chi_f^{\dagger}(v')
v^{\rho}v'^{\sigma}\chi_i^{\delta}(v),
\end{array}
\end{equation}
where $\chi_i(v)$, $\chi_f(v')$, $\chi_i^{\mu}(v)$ and $\chi_f^{\mu}(v')$
stand for the initial scalar, final scalar, initial vector and final vector
diquark fields in the baryons respectively.

It is noted that for the electromagnetic currents, the flavors do not change
at the two sides of the vertex, so that there are some extra symmetries as
used  by Georgi, Wise and Carone and the fact is expressed in the 
relations between
the coefficients. For example, in the case of the $\gamma-M-M^*$ vertex,
due to the current conservation (CVC), $f_1=-f_2$ is required, and from our
formulae, one can immediately prove that.\\

\noindent {\bf Appendix B: The amplitude square}
\vspace{0.3cm}

\noindent (1) $B_{(bb)_1}\rightarrow B_{(bc)_1}+l+\bar{\nu}$
\begin{eqnarray}
\Gamma_1 &=& \sum_{spins}|T_{1\lambda}l^{\lambda}|^2 \nonumber\\
&=& {8\over 9}|\xi(v\cdot v')|^2 Tr[A_{\lambda}A_{\lambda'}\bar u'u\bar uu'
+B^2\bar u'\gamma_{\lambda}u\bar u\gamma_{\lambda'}u'
+(A_{\lambda}B\bar u'u\bar u\gamma_{\lambda'}u'+C.T)\nonumber \\
&& +C^{\rho\sigma}C^{\rho'\sigma'}\epsilon_{\lambda\delta\rho\sigma}\epsilon_{
\lambda'\delta'\rho'\sigma'}\bar u'\gamma^{\delta}u\bar u\gamma^{\delta'}u'
-(C^{\rho\sigma}D^{\sigma'}\epsilon_{\lambda\delta\rho\sigma}\epsilon_{
\lambda'\delta'\rho'\sigma'}\bar u'\gamma^{\delta}u\bar u\gamma^{\rho'}
\gamma^{\delta'}u' \nonumber\\
&& +C.T)+D^{\sigma}D^{\sigma'}\epsilon_{\lambda\delta\rho\sigma}\epsilon_{
\lambda'\delta'\rho'\sigma'}\bar u'\gamma^{\delta}\gamma^{\rho}u\bar u
\gamma^{\rho'}\gamma^{\delta'}u'] \nonumber\\
&&\cdot (p_3^{\lambda}p_4^{\lambda'}+
p_3^{\lambda'}p_4^{\lambda}-(p_3\cdot p_4)g^{\lambda\lambda'}),
\end{eqnarray}
where
\begin{eqnarray}
A_{\lambda} &=& (-f_3v'_{\lambda}-f_4v_{\lambda})(2+v\cdot v')+(f_5v'_{\lambda}
+f_6v_{\lambda})(1-(v\cdot v')^2)- \nonumber\\
&& (f_7v'_{\lambda}+f_8v_{\lambda})(1+v\cdot v'),\\
B &=& (f_7+f_8)(1+v\cdot v'), \\
C^{\rho\sigma} &=& f_9v^{\rho}v'^{\sigma}-f_{10}v'^{\rho}v^{\sigma}, \\
D^{\sigma} &=& f_9v'^{\sigma}+f_{10}v^{\sigma}.
\end{eqnarray}

It is noted that  in $Tr[\rlap/p_3\gamma_{\lambda}(1-\gamma_5)\rlap /p_4
\gamma_{\lambda'}(1-\gamma_5)]=2Tr[\rlap/p_3\gamma_{\lambda}\rlap/p_4
\gamma_{\lambda'}- \rlap/p_3\gamma_{\lambda}\rlap/p_4\gamma_{\lambda'}
\gamma_5]$ there is a term
$\epsilon_{\alpha\lambda\alpha'\lambda'}p_3^{\alpha}p_4^{\alpha}$  which is
antisymmetric to an exchange of $p_3$ and $p_4$, so should  make
null contribution after integration over the final state phase space
as long as we omit the masses of leptons. The "C.T" means the conjugate
term, for example
$$(\bar u'\Gamma u\bar u\Gamma'u')^*=\bar u\gamma_0\Gamma^{\dagger}\gamma_0 u'
\bar u'\gamma_0\Gamma^{'\dagger}\gamma_0 u,$$
and the other parts would be changed accordingly. Later the terms concerning
the spinor-vector $\psi_{\mu}$ are also of their conjugate correspondence
which can be obtained in a rule similar to the spinors.

\noindent (2) $B_{(bb)_1}\rightarrow B_{(bc)_0}+l+\bar{\nu}$
\begin{eqnarray}
\Gamma_2 &=& \sum_{spins} |T_{2\lambda}l^{\lambda}|^2 \nonumber\\
&=& {8\over 3}|\xi(v\cdot v')|^2Tr[-A_{\lambda}A_{\lambda'}\bar u'\gamma_5u
\bar u\gamma_5 u'+(A_{\lambda}B\bar u'\gamma_5u\bar u\gamma_{\lambda'}
\gamma_5u'+C.T)\nonumber \\
&&+B^2\bar u\gamma_{\lambda}\gamma_5u\bar u\gamma_{\lambda'}\gamma_5u'
+C^{\rho\sigma}C^{\rho'\sigma'}\epsilon_{\lambda\delta\rho\sigma}
\epsilon_{\lambda'\delta'\rho'\sigma'}\bar u'\gamma^{\delta}\gamma_5u
\bar u\gamma^{\delta'}\gamma_5u']\nonumber \\
&&\cdot (p_3^{\lambda}p_4^{\lambda'}+
p_3^{\lambda'}p_4^{\lambda}-(p_3\cdot p_4)g^{\lambda\lambda'}),
\end{eqnarray}
where
\begin{eqnarray}
A_{\lambda} &=& (f_{16}v'_{\lambda}+f_{17}v_{\lambda})(1+v\cdot v')+f_{15}
v_{\lambda}, \\
B &=& f_{15}, \\
c^{\rho\sigma} &=& f_{18}v^{\rho}v'^{\sigma}.
\end{eqnarray}

\noindent (3) $B_{(bb)_1}\rightarrow B^*_{(bc)_1}+l+\bar{\nu}$
\begin{eqnarray}
\Gamma_3 &=& \sum_{spins} |T_{3\lambda}l^{\lambda}|^2 \nonumber \\
&=& {8\over 3}|\xi(v\cdot v')|^2Tr[-A_{\lambda\delta}A_{\lambda'\delta'}
\bar\Psi'^{\delta}\gamma_5u\bar u \gamma_5\Psi'^{\delta'}+(A_{\lambda\delta}
B_{\lambda'}
\bar\Psi'^{\delta}\gamma_5u\bar u\gamma_{\delta'}\gamma_5\Psi'^{\delta'}+C.T)
 \nonumber\\
&& +(A_{\lambda\delta}C_{\delta'}\bar\Psi'^{\delta}\gamma_5u\bar u
\gamma_{\lambda'}\gamma_5\Psi'^{\delta'}+C.T)
+B_{\lambda}B_{\lambda'}\bar\Psi'^{\delta}\gamma_{\delta}\gamma_5u\bar u
\gamma_{\delta'}\gamma_5\Psi'^{\delta'}\nonumber \\
&&+C_{\delta}C_{\delta'}\bar\Psi'^{\delta}\gamma_{\lambda}\gamma_5u
\bar u\gamma_{\lambda'}\gamma_5\Psi'^{\delta'} 
+(B_{\lambda}C_{\delta'}\bar\Psi'^{\delta}\gamma_{\delta}\gamma_5u
\bar u\gamma_{\lambda'}\gamma_5\Psi'^{\delta'}+C.T)\nonumber \\
&&+D^{\sigma}D^{\sigma'}
\epsilon_{\lambda\delta\rho\sigma} \epsilon_{\lambda'\delta'\rho'\sigma'}
\bar\Psi'^{\delta}\gamma^{\rho}\gamma_5u\bar u\gamma^{\rho'}\gamma_5
\Psi'^{\delta'} \nonumber \\
&& +(E^{\rho\sigma}D^{\sigma'}\epsilon_{\lambda\delta\rho\sigma}
\epsilon_{\lambda'\delta'\rho'\sigma'}\bar\Psi'^{\delta}\gamma_5u\bar u
\gamma^{\rho'}\gamma_5\Psi'^{\delta'}+C.T)\nonumber \\
&&-E^{\rho\sigma}E^{\rho'\sigma'}
\epsilon_{\lambda\delta\rho\sigma} \epsilon_{\lambda'\delta'\rho'\sigma'}
\bar\Psi'^{\delta}\gamma_5u\bar u\gamma_5\Psi'^{\delta'}] 
\cdot (p_3^{\lambda}p_4^{\lambda'}+
p_3^{\lambda'}p_4^{\lambda}-(p_3\cdot p_4)g^{\lambda\lambda'}),
\end{eqnarray} 
where
\begin{eqnarray}
A_{\lambda\delta} &=& f_3v'_{\lambda}v_{\delta}+f_4v_{\lambda}v_{\delta}
+(f_5v'_{\lambda}+f_6v_{\lambda})v_{\delta}(1+v\cdot v')+f_8v_{\delta}
v_{\lambda},\\
B_{\lambda} &=& f_3v'_{\lambda}+f_4v_{\lambda},\\
C_{\delta} &=& f_8v_{\delta},\\
D^{\sigma} &=& f_9v'^{\sigma}+f_{10}v^{\sigma},\\
E^{\rho\sigma} &=& f_9v^{\rho}v'^{\sigma}.
\end{eqnarray}

\noindent (4) $B^*_{(bb)_1}\rightarrow B^*_{(bc)_1}+l+\bar\nu$
\begin{eqnarray}
\Gamma_4 &=& \sum_{spins}|T_{4\lambda}l^{\lambda}|^2 \nonumber\\
&=& 8|\xi(v\cdot v')|^2Tr[A_{\lambda}A_{\lambda'}\bar\Psi'^{\delta}\Psi_{\delta}
\bar\Psi_{\delta'}\Psi'^{\delta'}+
 (A_{\lambda}B_{\nu'\mu'\lambda'}\bar\Psi'^{\delta}\Psi_{\delta}
\bar\Psi^{\nu'}\Psi'^{\mu'}+ C.T)\nonumber\\
&& (A_{\lambda}C_{\delta'}\bar\Psi'^{\delta}
\Psi_{\delta}\bar\Psi_{\lambda'}\Psi'^{\delta'}+C.T)
+ B_{\nu\mu\lambda}B_{\nu'\mu'\lambda'}\bar\Psi'^{\mu}\Psi^{\nu}
\bar\Psi^{\nu'}\Psi'^{\mu'} \nonumber \\
&& + (B_{\nu\mu\lambda}C_{\delta'}\bar\Psi'^{\mu}\Psi^{\nu}
\bar\Psi_{\lambda'}\Psi'^{\delta'}+C.T)+ C_{\delta}C_{\delta'}\bar\Psi'^{\delta}
\Psi_{\lambda}\bar\Psi_{\lambda'}\Psi'^{\delta'} \nonumber\\
&& +D^{\sigma}D^{\sigma'}\epsilon_{\lambda\delta\rho\sigma}\epsilon_{\lambda'
\delta'\rho'\sigma'}\bar\Psi'^{\delta}\Psi^{\rho}\bar\Psi^{\rho'}
\Psi'^{\delta'}] \nonumber \\
&& \cdot (p_3^{\lambda}p_4^{\lambda'}+p_3^{\lambda'}p_4^{\lambda}
-(p_3\cdot p_4)g^{\lambda\lambda'}),
\end{eqnarray}
where
\begin{eqnarray}
A_{\lambda} &=& f_3v'_{\lambda}+f_4v_{\lambda}, \\
B_{\nu\mu\lambda} &=& f_5v'_{\nu}v_{\mu}v'_{\lambda}+f_6v'_{\nu}v_{\mu}
v_{\lambda},\\
C_{\delta} &=& f_7v'_{\delta}+f_8v_{\delta}, \\
D^{\sigma} &=& f_9v'^{\sigma}+f_{10}v^{\sigma}.
\end{eqnarray}

\noindent (5) $ B^*_{(bb)_1}\rightarrow B_{(bc)_1}+l+\bar\nu$
\begin{eqnarray}
\Gamma_5 &=& \sum_{spins} |T_{5\lambda}l^{\lambda}|^2 \nonumber \\
&=& {8\over 3}|\xi(v\cdot v')|^2Tr[-A_{\lambda\alpha}A_{\lambda'\alpha'}
\bar u'\gamma_5\Psi^{\alpha}\bar\Psi^{\alpha'}\gamma_5u'-(A_{\lambda\alpha}B
\bar u'\gamma_5\Psi^{\alpha}\bar\Psi_{\lambda'}\gamma_5u'+C.T) \nonumber\\
&& +(A_{\lambda\alpha}C_{\lambda'}\bar u'\gamma_5\Psi^{\alpha}\bar\Psi^{\alpha'}
\gamma_{\alpha'}\gamma_5u'+C.T)-(A_{\lambda\alpha}D_{\alpha'}\bar u'\gamma_5
\Psi^{\alpha}\bar\Psi^{\alpha'}\gamma_{\lambda'}\gamma_5u'+C.T) \nonumber\\
&& -B^2\bar u'\gamma_5\Psi_{\lambda}\bar\Psi_{\lambda'}\gamma_5u'
+(BC_{\lambda'}\bar u'\gamma_5\Psi_{\lambda}\bar\Psi^{\alpha'}\gamma_{\alpha'}
\gamma_5u'+C.T)  \nonumber\\
&&-(BD_{\alpha'}\bar u'\gamma_5\Psi_{\lambda}\bar\Psi^{\alpha'}
\gamma_{\lambda'}\gamma_5u'+C.T) 
+C_{\lambda}C_{\lambda'}\bar u'\gamma_{\alpha}\gamma_5\Psi^{\alpha}\bar\Psi'^{
\alpha'}\gamma_{\alpha'}\gamma_5u' \nonumber\\
&&-(C_{\lambda}D_{\alpha'}\bar u'\gamma_{\alpha}
\gamma_5\Psi^{\alpha}\bar\Psi^{\alpha'}\gamma_{\lambda'}\gamma_5 u'+C.T) 
+D_{\alpha}D_{\alpha'}\bar u'\gamma_{\lambda}\gamma_5\Psi^{\alpha}
\bar\Psi^{\alpha'}\gamma_{\lambda'}\gamma_5u' \nonumber\\
&&-F^{\delta\sigma}F^{\delta'\sigma'}
\epsilon_{\lambda\delta\rho\sigma} \epsilon_{\lambda'\delta'\rho'\sigma'}
\bar u'\gamma_5\Psi^{\rho}\bar\Psi^{\rho'}\gamma_5u' \nonumber\\ 
&&-(G^{\sigma}F^{\delta'\sigma'}
\epsilon_{\lambda\delta\rho\sigma} \epsilon_{\lambda'\delta'\rho'\sigma'}
\bar u'\gamma^{\delta}\gamma_5\Psi^{\rho}\bar\Psi^{\rho'}\gamma_5u'
+C.T) \nonumber\\
&& +G^{\sigma}G^{\sigma'}
\epsilon_{\lambda\delta\rho\sigma} \epsilon_{\lambda'\delta'\rho'\sigma'}
\bar u'\gamma^{\delta}\gamma_5\Psi^{\rho}\bar\Psi^{\rho'}
\gamma^{\delta'}\gamma_5u'] \nonumber\\
&&\cdot (p_3^{\lambda}p_4^{\lambda'}+
p_3^{\lambda'}p_4^{\lambda}-(p_3\cdot p_4)g^{\lambda\lambda'}),
\end{eqnarray} 
where
\begin{eqnarray}
A_{\lambda\alpha} &=& f_3v'_{\lambda}v'_{\alpha}+f_4v_{\lambda}v'_{\alpha}
+f_5(1+v\cdot v')v'_{\alpha}v'_{\lambda} \nonumber\\
&& +f_6(1+v\cdot v')v_{\lambda}v'_{\alpha}+f_7v'_{\lambda}v'_{\alpha}, \\
B &=& f_8(1+v\cdot v'),\\
C &=& -f_3v'_{\lambda}-f_4v_{\lambda}, \\
D_{\alpha} &=& f_7v'_{\alpha}, \\
F^{\delta\sigma} &=& f_{10}v'^{\delta}v^{\sigma}, \\
G^{\sigma} &=& -f_9v'^{\sigma}-f_{10}v^{\sigma}.
\end{eqnarray}

\noindent (6) $B^*_{(bb)_1}\rightarrow B_{(bc)_0}+l+\bar\nu$
\begin{eqnarray}
\Gamma_6 &=& \sum_{spins}|T_{6\lambda}l^{\lambda}|^2 \nonumber\\
&=& 8|\xi(v\cdot v')|^2Tr[ A^2\bar u'\Psi_{\lambda}\bar\Psi_{\lambda'}u'
+(AB_{\delta'\lambda'}\bar u'\Psi_{\lambda}\bar\Psi^{\delta'}u'+C.T)\nonumber\\
&&+B_{\delta\lambda}B_{\delta'\lambda'}\bar u'\Psi^{\delta}\bar\Psi^{\delta'}u'
+ C^{\rho\sigma}C^{\rho'\sigma'}\epsilon_{\lambda\delta\rho\sigma}
\epsilon_{\lambda'\delta'\rho'\sigma'}\bar u'\Psi^{\delta}\bar\Psi^{\delta'}u'
] \nonumber \\
&& \cdot (p_3^{\lambda}p_4^{\lambda'}+p_3^{\lambda'}p_4^{\lambda}-
(p_3\cdot p_4)g^{\lambda\lambda'}),
\end{eqnarray}
where
\begin{eqnarray}
A &=& f_{15}, \\
B_{\delta\lambda} &=& f_{16}v'_{\delta}v'_{\lambda}+f_{17}v'_{\delta}
v_{\lambda}, \\
C^{\rho\sigma} &=& f_{18}v^{\rho}v'^{\sigma}.
\end{eqnarray}

\noindent (7) $B^*_{(bc)_1}\rightarrow B^*_{(cc)_1}+l+\bar\nu$
\begin{eqnarray}
\Gamma_7 &=& \sum_{spins}|T_{7\lambda}l^{\lambda}|^2 \nonumber\\
&=& 8|\xi(v\cdot v')|^2Tr[A_{\lambda}A_{\lambda'}\bar\Psi'^{\delta}\Psi_{\delta}
\bar\Psi_{\delta'}\Psi'^{\delta'}+
(A_{\lambda}B_{\nu'\mu'\lambda'}\bar\Psi'^{\delta}\Psi_{\delta}
\bar\Psi^{\nu'}\Psi'^{\mu'}+C.T) \nonumber\\
&& +(A_{\lambda}C_{\delta'}\bar\Psi'^{\delta}
\Psi_{\delta}\bar\Psi_{\lambda'}\Psi'^{\delta'}+C.T)
+ B_{\nu\mu\lambda}B_{\nu'\mu'\lambda'}\bar\Psi'^{\mu}\Psi^{\nu}
\bar\Psi^{\nu'}\Psi'^{\mu'} \nonumber \\
&& +(B_{\nu\mu\lambda}C_{\delta'}\bar\Psi'^{\mu}\Psi^{\nu}
\bar\Psi_{\lambda'}\Psi'^{\delta'}+C.T)+
C_{\delta}C_{\delta'}\bar\Psi'^{\delta}
\Psi_{\lambda}\bar\Psi_{\lambda'}\Psi'^{\delta'} \nonumber\\
&& +D^{\sigma}D^{\sigma'}\epsilon_{\lambda\delta\rho\sigma}\epsilon_{\lambda'
\delta'\rho'\sigma'}\bar\Psi'^{\delta}\Psi^{\rho}\bar\Psi^{\rho'}
\Psi'^{\delta'}](p_3^{\lambda}p_4^{\lambda'}+p_3^{\lambda'}p_4^{\lambda}
-(p_3\cdot p_4)g^{\lambda\lambda'}),
\end{eqnarray}
where
\begin{eqnarray}
A_{\lambda} &=& f_3v'_{\lambda}+f_4v_{\lambda}, \\
B_{\nu\mu\lambda} &=& f_5v'_{\nu}v_{\mu}v'_{\lambda}+f_6v'_{\nu}v_{\mu}
v_{\lambda},\\
C_{\delta} &=& f_7v'_{\delta}+f_8v_{\delta}, \\
D^{\sigma} &=& f_9v'^{\sigma}+f_{10}v^{\sigma}.
\end{eqnarray}

\noindent (8) $ B^*_{(bc)_1}\rightarrow B_{(cc)_1}+l+\bar\nu$
\begin{eqnarray}
\Gamma_8 &=& \sum_{spins} |T_{8\lambda}l^{\lambda}|^2 \nonumber \\
&=& {8\over 3}|\xi(v\cdot v')|^2Tr[-A_{\lambda\alpha}A_{\lambda'\alpha'}
\bar u'\gamma_5\Psi^{\alpha}\bar\Psi^{\alpha'}\gamma_5u'-(A_{\lambda\alpha}B
\bar u'\gamma_5\Psi^{\alpha}\bar\Psi_{\lambda'}\gamma_5u'+C.T) \nonumber\\
&& +(A_{\lambda\alpha}C_{\lambda'}\bar u'\gamma_5\Psi^{\alpha}\bar\Psi^{\alpha'}
\gamma_{\alpha'}\gamma_5u'+C.T)-(A_{\lambda\alpha}D_{\alpha'}\bar u'\gamma_5
\Psi^{\alpha}\bar\Psi^{\alpha'}\gamma_{\lambda'}\gamma_5u'+C.T) \nonumber\\
&& -B^2\bar u'\gamma_5\Psi_{\lambda}\bar\Psi_{\lambda'}\gamma_5u'
+(BC_{\lambda'}\bar u'\gamma_5\Psi_{\lambda}\bar\Psi^{\alpha'}\gamma_{\alpha'}
\gamma_5u'+C.T)\nonumber\\
&&-(BD_{\alpha'}\bar u'\gamma_5\Psi_{\lambda}\bar\Psi^{\alpha'}
\gamma_{\lambda'}\gamma_5u'+C.T) 
+C_{\lambda}C_{\lambda'}\bar u'\gamma_{\alpha}\gamma_5\Psi^{\alpha}\bar\Psi'^{
\alpha'}\gamma_{\alpha'}\gamma_5u'\nonumber\\
&&-(C_{\lambda}D_{\alpha'}\bar u'\gamma_{\alpha}
\gamma_5\Psi^{\alpha}\bar\Psi^{\alpha'}\gamma_{\lambda'}\gamma_5 u' +C.T)
+D_{\alpha}D_{\alpha'}\bar u'\gamma_{\lambda}\gamma_5\Psi^{\alpha}
\bar\Psi^{\alpha'}\gamma_{\lambda'}\gamma_5u'\nonumber\\
&&-F^{\delta\sigma}F^{\delta'\sigma'}
\epsilon_{\lambda\delta\rho\sigma} \epsilon_{\lambda'\delta'\rho'\sigma'}
\bar u'\gamma_5\Psi^{\rho}\bar\Psi^{\rho'}\gamma_5u' \nonumber\\
&&-(G^{\sigma}F^{\delta'\sigma'}
\epsilon_{\lambda\delta\rho\sigma} \epsilon_{\lambda'\delta'\rho'\sigma'}
\bar u'\gamma^{\delta}\gamma_5\Psi^{\rho}\bar\Psi^{\rho'}\gamma_5u'
+C.T) \nonumber\\
&& +G^{\sigma}G^{\sigma'}
\epsilon_{\lambda\delta\rho\sigma} \epsilon_{\lambda'\delta'\rho'\sigma'}
\bar u'\gamma^{\delta}\gamma_5\Psi^{\rho}\bar\Psi^{\rho'}
\gamma^{\delta'}\gamma_5u'] \nonumber \\
&& \cdot (p_3^{\lambda}p_4^{\lambda'}+
p_3^{\lambda'}p_4^{\lambda}-(p_3\cdot p_4)g^{\lambda\lambda'}),
\end{eqnarray} 
where
\begin{eqnarray}
A_{\lambda\alpha} &=& f_3v'_{\lambda}v'_{\alpha}+f_4v_{\lambda}v'_{\alpha}
+f_5(1+v\cdot v')v'_{\alpha}v'_{\lambda} \nonumber\\
&& +f_6(1+v\cdot v')v_{\lambda}v'_{\alpha}+f_7v'_{\lambda}v'_{\alpha}, \\
B &=& f_8(1+v\cdot v'),\\
C &=& -f_3v'_{\lambda}-f_4v_{\lambda}, \\
D_{\alpha} &=& f_7v'_{\alpha}, \\
F^{\delta\sigma} &=& f_{10}v'^{\delta}v^{\sigma}, \\
G^{\sigma} &=& -f_9v'^{\sigma}-f_{10}v^{\sigma}.
\end{eqnarray}

\noindent (9) $B_{(bc)_1}\rightarrow B^*_{(cc)_1}+l+\bar{\nu}$
\begin{eqnarray}
\Gamma_9 &=& \sum_{spins} |T_{9\lambda}l^{\lambda}|^2 \nonumber \\
&=& {8\over 3}|\xi(v\cdot v')|^2Tr[-A_{\lambda\delta}A_{\lambda'\delta'}
\bar\Psi'^{\delta}\gamma_5u\bar u \gamma_5\Psi'^{\delta'}+(A_{\lambda\delta}
B_{\lambda'}
\bar\Psi'^{\delta}\gamma_5u\bar u\gamma_{\delta'}\gamma_5\Psi'^{\delta'}
+C.T)  \nonumber\\
&& +(A_{\lambda\delta}C_{\delta'}\bar\Psi'^{\delta}\gamma_5u\bar u
\gamma_{\lambda'}\gamma_5\Psi'^{\delta'}+C.T)
+B_{\lambda}B_{\lambda'}\bar\Psi'^{\delta}\gamma_{\delta}\gamma_5u\bar u
\gamma_{\delta'}\gamma_5\Psi'^{\delta'}\nonumber\\
&&+C_{\delta}C_{\delta'}\bar\Psi'^{\delta}\gamma_{\lambda}\gamma_5u\bar u
\gamma_{\lambda'}\gamma_5\Psi'^{\delta'} 
+(B_{\lambda}C_{\delta'}\bar\Psi'^{\delta}\gamma_{\delta}\gamma_5u
\bar u\gamma_{\lambda'}\gamma_5\Psi'^{\delta'}+C.T)\nonumber\\
&&+D^{\sigma}D^{\sigma'}
\epsilon_{\lambda\delta\rho\sigma} \epsilon_{\lambda'\delta'\rho'\sigma'}
\bar\Psi'^{\delta}\gamma^{\rho}\gamma_5u\bar u\gamma^{\rho'}\gamma_5
\Psi'^{\delta'} \nonumber \\
&& +(E^{\rho\sigma}D^{\sigma'}\epsilon_{\lambda\delta\rho\sigma}
\epsilon_{\lambda'\delta'\rho'\sigma'}\bar\Psi'^{\delta}\gamma_5u\bar u
\gamma^{\rho'}\gamma_5\Psi'^{\delta'}+C.T)\nonumber\\
&&-E^{\rho\sigma}E^{\rho'\sigma'}
\epsilon_{\lambda\delta\rho\sigma} \epsilon_{\lambda'\delta'\rho'\sigma'}
\bar\Psi'^{\delta}\gamma_5u\bar u\gamma_5\Psi'^{\delta'}] \nonumber \\
&& \cdot (p_3^{\lambda}p_4^{\lambda'}+
p_3^{\lambda'}p_4^{\lambda}-(p_3\cdot p_4)g^{\lambda\lambda'}),
\end{eqnarray} 
where
\begin{eqnarray}
A_{\lambda\delta} &=& f_3v'_{\lambda}v_{\delta}+f_4v_{\lambda}v_{\delta}
+(f_5v'_{\lambda}+f_6v_{\lambda})v_{\delta}(1+v\cdot v')+f_8v_{\delta}
v_{\lambda},\\
B_{\lambda} &=& +f_3v'_{\lambda}+f_4v_{\lambda},\\
C_{\delta} &=& f_8v_{\delta},\\
D^{\sigma} &=& f_9v'^{\sigma}+f_{10}v^{\sigma},\\
E^{\rho\sigma} &=& f_9v^{\rho}v'^{\sigma}.
\end{eqnarray}

\noindent (10) $B_{(bc)_1}\rightarrow B_{(cc)_1}+l+\bar{\nu}$
\begin{eqnarray}
\Gamma_{10} &=& \sum_{spins}|T_{10\lambda}l^{\lambda}|^2 \nonumber\\
&=& {8\over 9}|\xi(v\cdot v')|^2 Tr[A_{\lambda}A_{\lambda'}\bar u'u\bar uu'
+B^2\bar u'\gamma_{\lambda}u\bar u\gamma_{\lambda'}u'
+(A_{\lambda}B\bar u'u\bar u\gamma_{\lambda'}u'+C.T) \nonumber \\
&& C^{\rho\sigma}C^{\rho'\sigma'}\epsilon_{\lambda\delta\rho\sigma}\epsilon_{
\lambda'\delta'\rho'\sigma'}\bar u'\gamma^{\delta}u\bar u\gamma^{\delta'}u'
-(C^{\rho\sigma}D^{\sigma'}\epsilon_{\lambda\delta\rho\sigma}\epsilon_{
\lambda'\delta'\rho'\sigma'}\bar u'\gamma^{\delta}u\bar u\gamma^{\rho'}
\gamma^{\delta'}u' \nonumber\\
&& +C.T)+D^{\sigma}D^{\sigma'}\epsilon_{\lambda\delta\rho\sigma}\epsilon_{
\lambda'\delta'\rho'\sigma'}\bar u'\gamma^{\delta}\gamma^{\rho}u\bar u
\gamma^{\rho'}\gamma^{\delta'}u'] \nonumber\\
&& \cdot (p_3^{\lambda}p_4^{\lambda'}+
p_3^{\lambda'}p_4^{\lambda}-(p_3\cdot p_4)g^{\lambda\lambda'}),
\end{eqnarray}
where
\begin{eqnarray}
A_{\lambda} &=& (-f_3v'_{\lambda}-f_4v_{\lambda})(2+v\cdot v')+(f_5v'_{\lambda}
+f_6v_{\lambda})(1-(v\cdot v')^2) \nonumber\\
&& -(f_7v'_{\lambda}+f_8v_{\lambda})(1+v\cdot v'),\\
B &=& (f_7+f_8)(1+v\cdot v'), \\
C^{\rho\sigma} &=& f_9v^{\rho}v'^{\sigma}-f_{10}v'^{\rho}v^{\sigma}, \\
D^{\sigma} &=& f_9v'^{\sigma}+f_{10}v^{\sigma}.
\end{eqnarray}

\noindent (11) $B_{(bc)_0}\rightarrow B^*_{(cc)_1}+l+\bar\nu$
\begin{eqnarray}
\Gamma_{11} &=& \sum_{spins}|T_{11\lambda}l^{\lambda}|^2 \nonumber \\
&=& 8|\xi(v\cdot v')|^2Tr[\{ A^2\bar\Psi'_{\lambda}u\bar u\Psi'_{\lambda'}
+(AB_{\delta'\lambda'}\bar\Psi'_{\lambda}u\bar u\Psi'^{\delta'}+C.T)
\nonumber \\
&&+B_{\delta\lambda}B_{\delta'\lambda'}\bar\Psi'^{\delta}u\bar u\Psi'^{\delta'}
+C^{\rho\sigma}C^{\rho'\sigma'}\epsilon_{\lambda\delta\rho\sigma}
\epsilon_{\lambda'\delta'\rho'\sigma'}\bar\Psi'^{\delta}u\bar u\Psi'^{\delta'}
\}] \nonumber \\
&& \cdot (p_3^{\lambda}p_4^{\lambda'}+
p_3^{\lambda'}p_4^{\lambda}-(p_3\cdot p_4)g^{\lambda\lambda'}),
\end{eqnarray}
where
\begin{eqnarray}
A &=& f_{11}, \\
B_{\delta\lambda} &=& f_{12}v_{\delta}v'_{\lambda}+f_{13}v_{\delta}v_{\lambda}, \\
C^{\rho\sigma} &=& f_{14}v'^{\rho}v^{\sigma}.
\end{eqnarray}

\noindent (12) $ B_{(bc)_0}\rightarrow B_{(cc)_1}+l+\bar\nu$
\begin{eqnarray}
\Gamma_{12} &=& \sum_{spins}|T_{12\lambda}l^{\lambda}|^2 \nonumber \\
&=& {8\over 3}|\xi(v\cdot v')|^2 Tr[
-A_{\lambda}A_{\lambda'}\bar u'\gamma_5u\bar u\gamma_5 u'
+(A_{\lambda}B\bar u'\gamma_5u\bar u\gamma_{\lambda'}\gamma_5u'+C.T)
\nonumber \\
&&+B^2\bar u'\gamma_{\lambda}\gamma_5u\bar u\gamma_{\lambda'}\gamma_5u'
+C^{\rho\sigma}C^{\rho'\sigma'}\epsilon_{\lambda\delta\rho\sigma}
\epsilon_{\lambda'\delta'\rho'\sigma'}\bar u'\gamma^{\delta}\gamma_5u
\bar u\gamma^{\delta'}\gamma_5u']\nonumber \\
&&\cdot (p_3^{\lambda}p_4^{\lambda'}+
p_3^{\lambda'}p_4^{\lambda}-(p_3\cdot p_4)g^{\lambda\lambda'}),
\end{eqnarray}
where
\begin{eqnarray}
A_{\lambda} &=& f_{11}v'_{\lambda}+f_{12}(1+v\cdot v')v'_{\lambda}+
f_{13}(1+v\cdot v')v_{\lambda}, \\
B &=& -f_{11}, \\
C^{\rho\sigma} &=& f_{14}v'^{\rho}v^{\sigma}.
\end{eqnarray}

\end{document}